\documentclass[11pt,dvips]{article}

\usepackage{amsmath,epsfig,ulem}
\usepackage{rotating}
\usepackage{cite}
\usepackage{hhline}
\usepackage{array}
\usepackage{amssymb}
\usepackage{color}
\usepackage{dsfont}
\usepackage{axodraw}

\begin{document}
\tolerance=100000

\newcommand{\imag}{\Im {\rm m}}
\newcommand{\real}{\Re {\rm e}}

\def\tablename{\bf Table}%
\def\figurename{\bf Figure}%

\newcommand{\sts}{\scriptstyle}
\newcommand{\ngs}{\!\!\!\!\!\!}
\newcommand{\rb}[2]{\raisebox{#1}[-#1]{#2}}
\newcommand{\CP}{${\cal CP}$~}
\newcommand{\sbomu}{\frac{\sin 2 \beta}{2 \mu}}
\newcommand{\kmol}{\frac{\kappa \mu}{\lambda}}
\newcommand{\s}{\\ \vspace*{-3.5mm}}
\newcommand{\lsim}{\raisebox{-0.13cm}{~\shortstack{$<$\\[-0.07cm] $\sim$}}~}
\newcommand{\gsim}{\raisebox{-0.13cm}{~\shortstack{$>$\\[-0.07cm] $\sim$}}~}
\newcommand{\kr}{\color{red}}
\begin{titlepage}

\begin{flushright}
DESY 06-237\\
KEK-TH-1124\\
KIAS-P06063\\
\today
\end{flushright}

\vspace{0.6cm}

\begin{center}
{\Large \bf $\tau$ Polarization in SUSY Cascade Decays}\\[1cm]
{\large S.Y. Choi$^{1,2}$, K. Hagiwara$^{3}$, Y.G. Kim$^{4}$,
        K. Mawatari$^{5}$
        and P.M. Zerwas$^{2,3}$}\\[0.8cm]
{\it $^1$ Physics Department and RIPC, Chonbuk University, Jeonju 561-756,
          Korea\\
     $^2$ Deutsches Elektronen-Synchrotron DESY, D-22603 Hamburg, Germany\\
     $^3$ Theory Division, KEK, Tsukuba, Ibaraki 305-0801, Japan\\
     $^4$ ARCSEC, Sejong University, Seoul 143-747, Korea\\
     $^5$ School of Physics, Korea Institute for Advanced Study, Seoul 130-722, Korea
     }
\end{center}

\renewcommand{\thefootnote}{\fnsymbol{footnote}}
\vspace{1.8cm}

\begin{abstract}
\noindent
$\tau$ leptons emitted in cascade decays of supersymmetric
particles are polarized. The polarization may be exploited
to determine spin and mixing properties
of the neutralinos and stau particles involved.
\end{abstract}

\end{titlepage}

\newpage
\renewcommand{\thefootnote}{\fnsymbol{footnote}}


\noindent
{\bf 1.)} A rich source of information on supersymmetric particles
and the structure of the underlying theory can eventually be provided at
LHC \cite{Armstrong:1994it,unknown:1994pu} by cascade decays. After squarks
and gluinos are copiously produced
in high energy proton-proton collisions \cite{Beenakker:1996ch},
they cascade down, generally
in several steps, to the lightest supersymmetric particle, thereby generating
intermediate non-colored supersymmetric states like charginos/neutralinos
and sleptons \cite{Bachacou:1999zb} which are difficult to produce
otherwise at a hadron collider.\s

Much attention has been paid in the recent past to the SPS1a cascade
\cite{Aguilar-Saavedra:2005pw}
\begin{eqnarray}
\tilde{q}\ \ \to\ \ q\tilde{\chi}^0_2\ \ \to\ \
q\ell\tilde{\ell}\ \ \to \ \ q\ell\ell\tilde{\chi}^0_1
\label{eq:susy_cascade_chain}
\end{eqnarray}
which gives rise to a well-populated ensemble of neutralinos and $R$-type
sleptons with ratios of 30\%, 10\% and 100\% in the first, second and third
branching, respectively. Analyzing the visible $q \ell \ell$ final state in
various combinations of leptons and quark jet, the cascade has served to
study the precision with which the masses of supersymmetric particles can
be measured at LHC, for a summary see Ref.$\,$\cite{Weiglein}.
In addition, invariant mass distributions have been shown sensitive to the spin
of the particles involved \cite{Barr:2004ze,Goto:2004cp,Smillie:2005ar,
Wang:2006hk}, shedding light on the very nature of the new
particles observed in the cascade and on the underlying physics scenario.\s

So far, cascades have primarily been studied involving first and second
generation leptons/sleptons. In this brief note we will explore how the
polarization of $\tau$ leptons can be exploited to study $R/L$ chirality
and mixing effects in both the $\tilde{\tau}$ and the neutralino
sectors.{\footnote{For a discussion of polarization effects in single $\tau$
decays see Ref.\cite{MMN}.}}
As expected, it turns out that measuring the correlation of the $\tau$
polarizations provides
an excellent instrument to analyze these effects.\s

As polarization analyzer we will use single pion decays of the $\tau$'s.
At high energies the mass of the $\tau$ leptons can be neglected and the
fragmentation functions are linear in the fraction $z$ of
the energy transferred from the polarized $\tau$'s to the
$\pi$'s \cite{Bullock:1992yt}:
\begin{eqnarray}
(\tau_R)^\pm\, \to\ \ \stackrel{(-)}{\nu_\tau}\pi^\pm \;\; : \;\;\;\; F_R
  &=& 2 z
\label{eq:tau_to_pi_R}\\
(\tau_L)^\pm\, \to\ \ \stackrel{(-)}{\nu_\tau}\pi^\pm
                          \;\; :{\hspace*{.5mm}} \;\;\;\; F_L &=& 2 (1-z)
\label{eq:tau_to_pi_L}
\end{eqnarray}
In the relativistic limit, helicity and chirality are of equal and opposite
sign for $\tau^-$ leptons and $\tau^+$ anti-leptons, respectively. For notational
convenience we characterize the $\tau$ states by chirality. \s

This note should serve only as an exploratory theoretical study. Experimental
simulations will include other $\tau$ decay final states in addition to pions,
{\it e.g.} $\rho$'s and $a_1$'s. The $\rho$-meson mode is expected to contribute to
the $\tau$-spin correlation measurement even if the $\pi^\pm$ and $\pi^0$ energies
are not measured separately. In this case the $\tau$ polarization analysis power
of the $\rho$ channel is $\kappa_\rho
=(m^2_\tau-2m^2_\rho)/(m^2_\tau+2 m^2_\rho)\sim 1/2$ in contrast to $\kappa_\pi=1$,
but its larger branching fraction of {\rm B}$_\rho\approx 0.25$, {\it versus}
{\rm B}$_\pi \approx 0.11$, more than compensates for the reduced analysis power.
Moreover, in actual experiments it should be possible to measure the $\pi^\pm$
energy and the $\gamma$ energies of the $\pi^0$'s, all emitted along the parent
$\tau$-momentum direction at high energies. Significant improvement of the $\tau$
analysis power is therefore expected from the $\rho$ and $a_1$ modes by determining
the momentum fraction of $\pi^\pm$ in the collinear limit
of their decays \cite{Bullock:1992yt}.
For each mode, cuts and efficiencies for $\tau$ identification must be included
to arrive finally at realistic error estimates. The large size of the polarization
effects predicted on the theoretical basis, and exemplified quantitatively by the
pion channel, should guarantee their survival in realistic experimental environments,
and we expect that they can be exploited experimentally in practice.\s\s


%
\begin{figure}[ht!]
{\color{black}
\begin{center}
\begin{minipage}{8.7cm}
\begin{picture}(180,250)(-10,0)
\Text(0,100)[c]{(b)}
\Text(15,50)[r]{$q$}
\LongArrow(70,50)(20,50)
\Line(73,50)(120,50)
\Text(70,50)[c]{\Large $\otimes$}
\Photon(74,50)(120,50){2}{7}
\Text(120,50)[c]{\large $\bullet$}
\Text(85,68)[c]{\color{blue} $\theta_{\tau_n}$}
\LongArrowArc(120,50)(25,125,173)
\LongArrow(120,50)(95,90)
\Text(93,99)[]{$\tau_{(n)}$}
\Line(121,50)(136,26)
\Line(119,49)(134,25)
\Text(135,25)[c]{\large $\bullet$}
\Text(70,35)[c]{\color{red} $\tilde{q}$}
\Text(134,53)[c]{\color{red} $\tilde{\chi}^0_2$}
\Text(143,16)[c]{\color{red} $\tilde{\tau}_1$}
\Text(94,20)[c]{\color{blue} $\theta_{\tau_f}, \phi_{\tau_f}$}
\LongArrowArc(135,25)(20,125,214)
\Line(135,26)(165,50)
\Photon(135,26)(165,50){2}{7}
\Text(170,53)[l]{\color{red} $\tilde{\chi}^0_1$}
\LongArrow(135,26)(105,2)
\Text(100,-7)[c]{$\tau_{(f)}$}
\Text(0,250)[c]{(a)}
\Line(0,169)(50,169)
\Line(0,171)(50,171)
\Text(50,170)[c]{\large $\bullet$}
\Text(25,155)[c]{\color{red} $\tilde{q}^p_\alpha$}
\Text(45,200)[c]{$q^p_\alpha$}
\Line(50,170)(60,210)
\Line(50,170)(100,170)
\Photon(50,170)(100,170){2}{7}
\Text(100,170)[c]{\large $\bullet$}
\Text(75,155)[c]{\color{red} $\tilde{\chi}^0_j$}
\Text(95,200)[c]{$\tau^a_\beta$}
\Text(105,240)[c]{$\pi^a$}
\Line(100,170)(110,210)
\Text(110,210)[c]{\large $\bullet$}
\DashLine(110,210)(120,250){2}
\Line(100,169)(150,169)
\Line(100,171)(150,171)
\Text(125,155)[c]{\color{red} $\tilde{\tau}^{-a}_k$}
\Text(145,200)[c]{$\tau^{-a}_\gamma$}
\Text(155,240)[c]{$\pi^{-a}$}
\Line(150,170)(160,210)
\Text(150,170)[c]{\large $\bullet$}
\Text(160,210)[c]{\large $\bullet$}
\DashLine(160,210)(170,250){2}
\Line(150,170)(200,170)
\Photon(150,170)(200,170){2}{7}
\Text(175,155)[c]{\color{red} $\tilde{\chi}^0_1$}
\end{picture}
\end{minipage}
\begin{minipage}{6.9cm}
\mbox{ }\\[0.2cm]
\includegraphics[height=9.9cm,width=7.8cm,angle=0]{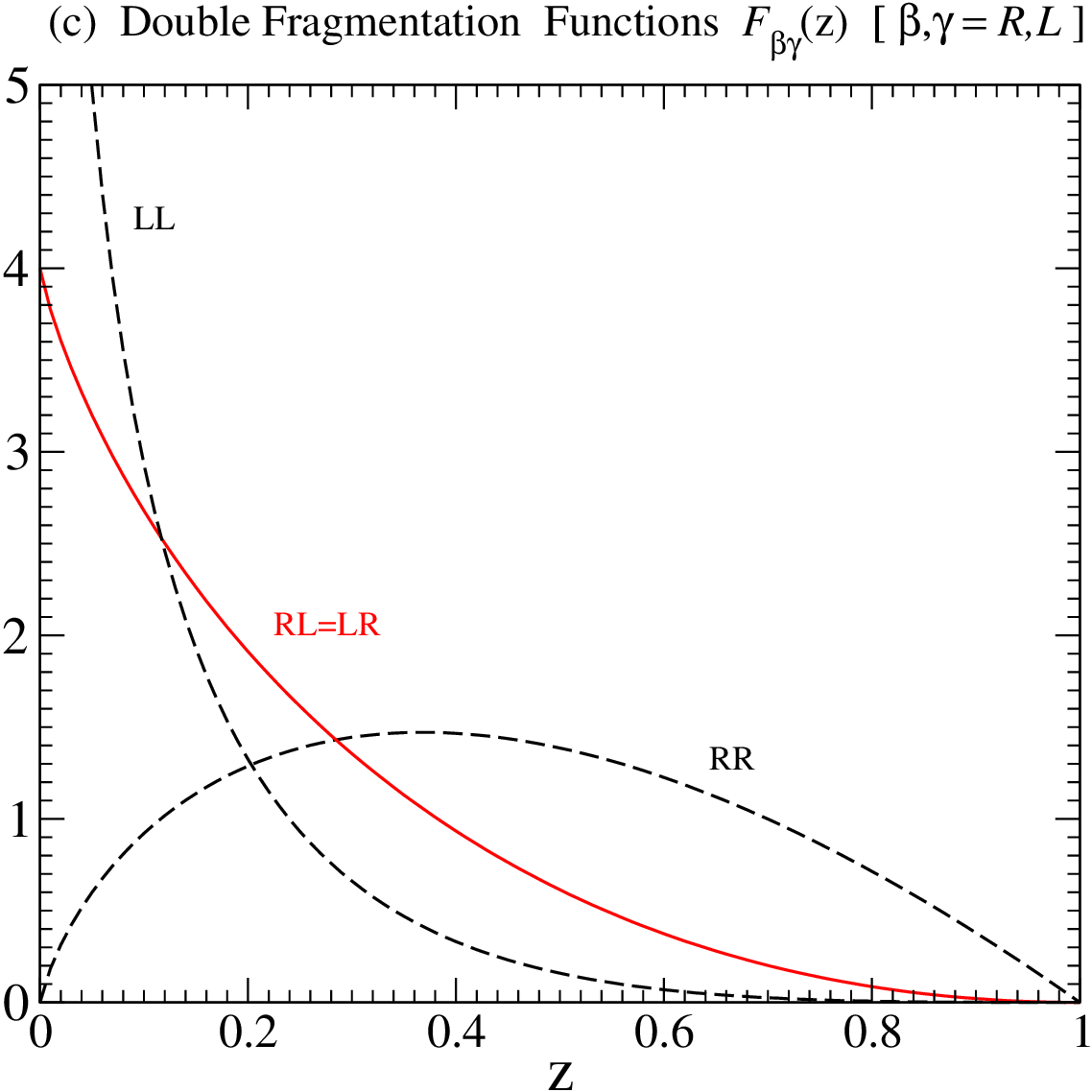}
\end{minipage}
\hskip 1.2cm \mbox{ }
\end{center}
}
\vskip 0.2cm
\caption{\it (a) The general structure of the quantum numbers of the particles
        involved in the cascade (\ref{eq:susy_cascade_chain});
        (b) the angular configuration of the particles in the
        squark/$\tilde{\chi}^0_2$/$\tilde{\tau}_1$ rest frames; and (c) the
        spectrum of the double $\tau_\beta \tau_\gamma \to \pi \pi$ fragmentation functions
        $F_{\beta\gamma}$ ($\beta,\gamma=R,L$) with $z=m^2_{\pi\pi}/m^2_{\tau\tau}$. }
\label{fig:diagram_fragmentation}
\end{figure}
\vskip 0.5cm

\newpage
\noindent
{\bf 2.)} The distribution of the visible final state particles in the squark
cascade can be cast, for massless quark $q$ and no squark $\tilde{q}$ mixing,
in the general form \cite{Goto:2004cp}:
\begin{eqnarray}
\frac{1}{\Gamma_{\tilde{q}_\alpha}}\frac{d\Gamma^{pa;jk}_{\alpha\beta\gamma}}{
  d\cos\theta_{\tau_n}d\cos\theta_{\tau_f}d\phi_{\tau_f}}
  = \frac{1}{8\pi}
      {\rm B}(\tilde{q}_\alpha\to q_\alpha\tilde{\chi}^0_j)\,
      {\rm B}(\tilde{\chi}^0_j\to \tau_\beta \tilde{\tau}_k)\,
      {\rm B}(\tilde{\tau}_k\to\tau_\gamma\tilde{\chi}^0_1)\,
      \left[ 1+(pa)(\alpha\beta) \cos\theta_{\tau_n} \right]
\label{eq:susy_distribution}
\end{eqnarray}
with $j=2$ and $k=1$ for the squark chain of Eq.$\,$(\ref{eq:susy_cascade_chain}).
The structure of the quantum numbers in the cascade is depicted in
Fig.$\,$\ref{fig:diagram_fragmentation}(a)
while the configuration of the particles in the squark/$\tilde{\chi}^0_2$/$\tilde{\tau}_1$
rest frames
is shown in Fig.$\,$\ref{fig:diagram_fragmentation}(b). For clarity the definitions
are summarized in the following table:
\begin{eqnarray}
\begin{array}{llll}
p =\pm :   & \mbox{particle/anti-particle}\hskip 2.3cm { }
           & \alpha=\pm :  & \mbox{$\tilde{q}$ and $q$ $R/L$ chirality} \\
a =\pm :   & \mbox{$\tau$ and $\pi$ charge}
           & \beta=\pm :   & \mbox{near $\tau_{n}$ $R/L$ chirality}\\
j=2,3,4:   & \mbox{neutralino mass index}
           & \gamma=\pm :  & \mbox{far $\tau_{f}$ $R/L$ chirality} \\
k=1,2 :    & \mbox{$\tilde{\tau}$ mass index}
           & { } & { }
\end{array}
\end{eqnarray}
Near $(n)$ and far $(f)$ indices denote $\tau$ and $\pi$ particles
emitted in $\tilde{\chi}^0_j$ and $\tilde{\tau}_k$ decays, respectively. \s

The quark/squark/neutralino $q_\alpha\tilde{q}_\alpha\tilde{\chi}^0_j$
and the tau/stau/neutralino $\tau_\beta\tilde{\tau}_k\tilde{\chi}^0_j$
vertices are given by the proper current couplings and the neutralino and
stau mixing matrix elements,
\begin{eqnarray}
\langle \tilde{\chi}^0_j| \tilde{q}_\alpha | f_\alpha\rangle
\, =\, i g\, A^q_{\alpha \alpha j}  \quad {\rm and } \quad
\langle \tilde{\chi}^0_j| \tilde{\tau}_k | \tau_\beta \rangle
\, =\, i g\, A^\tau_{\beta kj} \ \ [\gamma \;\; {\rm correspondingly}]
\label{eq:interaction_vertex}
\end{eqnarray}
with the explicit form of the couplings $A^q_{\alpha\alpha j}$ and
$A^\tau_{R k j}/A^\tau_{Lkj}$
\begin{eqnarray}
&& A^q_{LLj} = -\sqrt{2} [T_3^q N^*_{j2}+(e_q-T^q_3) N^*_{j1} t_W]\\
&& A^q_{RRj} = +\sqrt{2} e_q N_{j1} t_W\\
&& A^\tau_{L k j} = -h_\tau N^*_{j3} U_{\tilde{\tau}_{k2}}
                 +\frac{1}{\sqrt{2}} (N^*_{j2}+N^*_{j1} t_W)
                  U_{\tilde{\tau}_{k1}}
\label{eq:tau_L_coupling}\\
&& A^\tau_{R k j} = -h_\tau N_{j3} U_{\tilde{\tau}_{k1}}
                -\sqrt{2} N_{j1} t_W U_{\tilde{\tau}_{k2}}
\label{eq:tau_R_coupling}
\end{eqnarray}
in terms of the $4\times 4$ neutralino mixing matrix $N$ in the standard
gaugino/higgsino basis \cite{Choi:2001ww} and the $2\times 2$ stau mixing
matrix $U_{\tilde{\tau}}$ in the $L/R$ basis \cite{Boos:2003vf}.
Here, $T^q_3=\pm 1/2$ and $e_q=2/3,-1/3$ are the SU(2) doublet quark isospin
and electric charge, $t_W=\tan\theta_W$ and $h_\tau=m_\tau/\sqrt{2}m_W\cos\beta$.
The distribution (\ref{eq:susy_distribution}) depends only
on the ``near
$\tau$'' angle $\theta_{\tau_n}$; this is a consequence of the scalar character
of the intermediate stau state that erases all angular correlations.\s

The angles in the cascade Fig.$\,$\ref{fig:diagram_fragmentation}(b) are related
to the invariant masses \cite{Smillie:2005ar},
\begin{eqnarray}
&& { } \hskip -1.cm m^2_{\tau\tau}\,\, =\,
     \frac{1}{2}\left(1-\cos\theta_{\tau_f}\right)
               \hskip 3.8cm
   max\, M^2_{\tau\tau}\,\, =\, m^2_{\tilde{\chi}^0_j}
      (1-m^2_{\tilde{\tau}_k}/m^2_{\tilde{\chi}^0_j})
      (1-m^2_{\tilde{\chi}^0_1}/m^2_{\tilde{\tau}_k}) \\[1mm]
&& { } \hskip -1.cm m^2_{q\tau_{n}} = \,
     \frac{1}{2}\left(1-\cos\theta_{\tau_n}\right)
     \hskip 3.8cm
 max\, M^2_{q\tau_{n}}\, =\, m^2_{\tilde{q}_\alpha}
      (1-m^2_{\tilde{\chi}^0_j}/m^2_{\tilde{q}_\alpha})
      (1-m^2_{\tilde{\tau}_k}/m^2_{\tilde{\chi}^0_j})\\
&& { }\hskip -1.cm m^2_{q\tau_{f}} =\,\frac{1}{4}(1+c_n)(1-c_f)
   -\frac{r_{jk}}{2} s_n
   s_f \cos\phi_{\tau_f}
  { }\hskip 0.35cm { }
   max\, M^2_{q\tau_{f}}\, =\, m^2_{\tilde{q}_\alpha}
      (1-m^2_{\tilde{\chi}^0_j}/m^2_{\tilde{q}_\alpha})
      (1-m^2_{\tilde{\chi}^0_1}/m^2_{\tilde{\tau}_k}) \\
&& { }\hskip 0.25cm  +\frac{r^2_{jk}}{4} (1-c_n)(1+c_f)\nonumber
\end{eqnarray}
where $r_{jk} =m_{\tilde{\tau}_k}/m_{\tilde{\chi}^0_j}$ and abbreviations
$c_n = \cos\theta_{\tau_n}, c_f=\cos\theta_{\tau_f}$ {\it etc} are introduced.
It proves useful to define the invariant masses, $m^2_{\tau\tau}$,
$m^2_{q\tau_{n}}$ and $m^2_{q\tau_{f}}$, in units of their maximum values,
$max\, M^2_{\tau\tau}$, $max\, M^2_{q\tau_{n}}$ and
$max\, M^2_{q\tau_{f}}$; the invariant masses $m^2_{\pi\pi}$ etc are scaled
analogously. \s

Pion distributions, summed over near and far particles, are predicted by folding
the original single $\tau$ and double $\tau \tau$ distributions,
$d\Gamma_\beta / dm^2_{q\tau}$
and $d\Gamma_{\beta\gamma} / dm^2_{\tau\tau}$, with the single and double
fragmentation functions $F_\beta$ and $F_{\beta\gamma}$, where the indices
$\beta,\gamma$ denote the chirality indices $R/L$ of the $\tau$ leptons.
Based on standard techniques, the following relations can easily be derived
for [$q \pi$] and $[\pi \pi]$ distributions:
\begin{eqnarray}
\frac{d\Gamma}{dm^2_{q\pi}}
  & =&  \int^1_{m^2_{q\pi}} \frac{dm^2_{q\tau}}{m^2_{q\tau}}\,
   \frac{d\Gamma_\beta}{dm^2_{q\tau}}\,\,
   F_\beta\left(\frac{m^2_{q\pi}}{m^2_{q\tau}}\right) \\[3mm]
\frac{d\Gamma}{dm^2_{\pi\pi}}
  & =&  \int^1_{m^2_{\pi\pi}} \frac{dm^2_{\tau\tau}}{m^2_{\tau\tau}}\,
   \frac{d\Gamma_{\beta\gamma}}{dm^2_{\tau\tau}}\,\,
   F_{\beta\gamma}\left(\frac{m^2_{\pi\pi}}{m^2_{\tau\tau}}\right)
\end{eqnarray}
The single and double distributions
$d\Gamma_\beta / dm^2_{q\tau}$ and
$d\Gamma_{\beta\gamma} / dm^2_{\tau\tau}$,
can be derived from Eq.$\,$(\ref{eq:susy_distribution}) by integration.
The single $\tau_{\beta} \to \pi$ fragmentation function,
{\it cf.} Eqs.$\,$(\ref{eq:tau_to_pi_R}) and (\ref{eq:tau_to_pi_L}) with
$z=m^2_{q\pi}/m^2_{q\tau}$, can be summarized as
\begin{eqnarray}
\mbox{ }\hskip -1.8cm F_\beta(z) = 1 +\beta\, (2z-1)
\end{eqnarray}
while the double $\tau_{\beta} \tau_\gamma \to \pi \pi$
fragmentation functions, with $z=m^2_{\pi\pi}/m^2_{\tau\tau}$,
are given by
\begin{eqnarray}
&& F_{RR} (z) \,=\, 4\, z\, \log \frac{1}{z} \\
&& F_{RL} (z) \,=\, F_{LR} (z) \, =\, 4\left[1-z-z\,\log\frac{1}{z}\right]  \\
&& F_{LL} (z) \,=\, 4\left[(1+z)\,\log\frac{1}{z}+2z-2\right]
\end{eqnarray}
The shape of the distributions $F_{\beta\gamma}(z)$ ($\beta,\gamma=R,L$) is
presented in Fig.$\,$\ref{fig:diagram_fragmentation}(c).
All distributions, normalized to unity, are finite except $F_{LL}$
which is logarithmically divergent for $z \to 0$.\s

\begin{figure}[ht!]
\begin{center}
\includegraphics[height=9cm,width=14cm,angle=0]{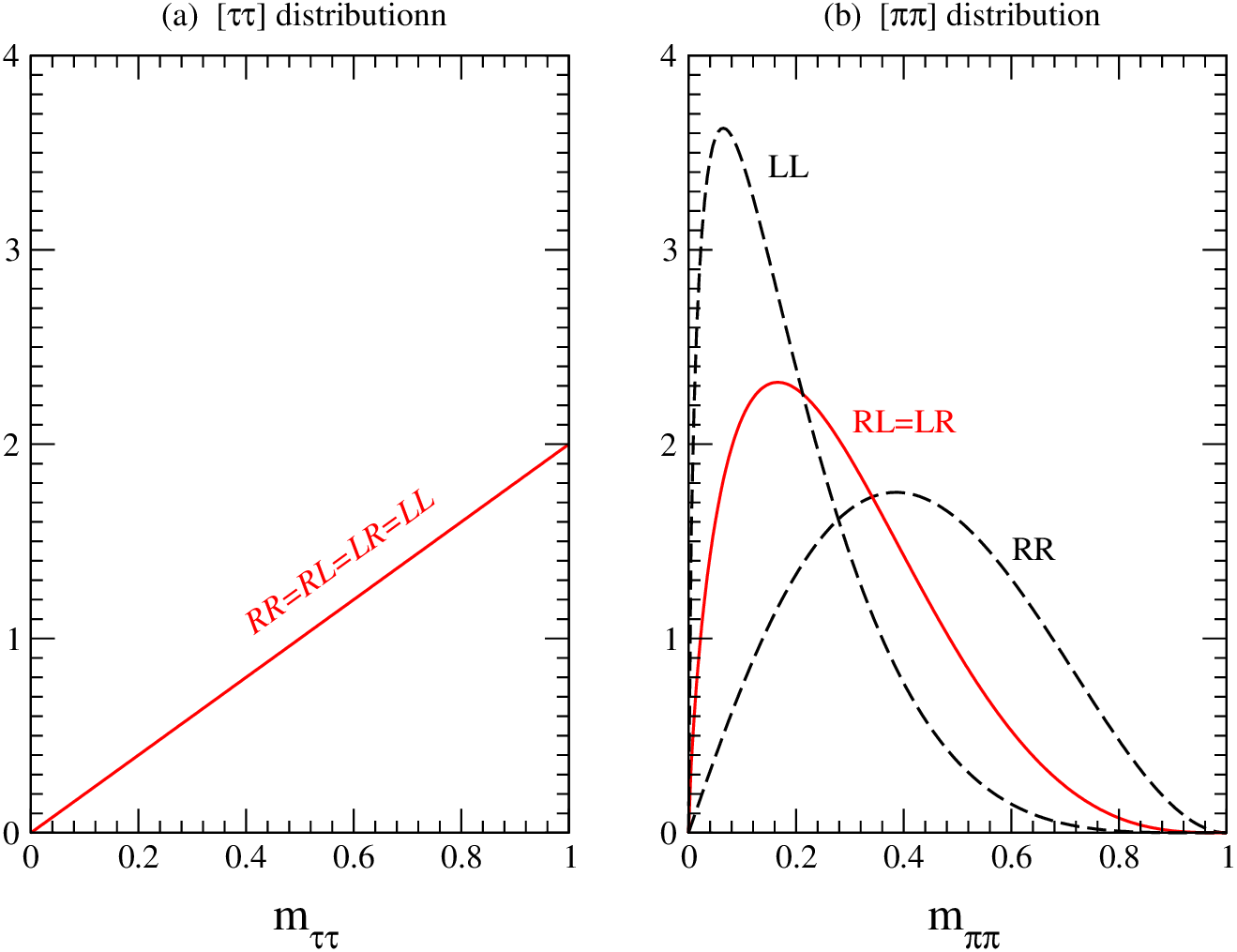} 
\end{center}
\vskip -0.5cm
\caption{\it The ``normalized" invariant mass distributions of (a) $\tau\tau$
    and (b) $\pi\pi$ pairings in the $\tilde{\chi}^0_2$ decays of the cascade
    (\ref{eq:susy_cascade_chain}). The indices denote the chiralities of
    the near and far tau leptons.}
\label{fig:tautau_vs_pipi}
\end{figure}

The great potential of polarization measurements for determining spin and
mixing phenomena is demonstrated in Figs.$\,$\ref{fig:tautau_vs_pipi}(a/b),
displaying the invariant mass distribution of (a) $\tau \tau$  and (b) $\pi \pi$
pairings in the $\tilde{\chi}^0_2$ decays of the cascade
(\ref{eq:susy_cascade_chain}). While the lepton-lepton
invariant mass does not depend on the chirality indices of the near and far tau
leptons $\tau_{n}$ and $\tau_{f}$, the shape of the pion distribution depends
strongly on the indices, as expected.
This is also reflected in the expectation values of the invariant masses:
\begin{eqnarray}
\langle m^2_{\pi\pi} \rangle \, =\,
\left\{\begin{array}{l}
        4/18\\[1mm]
        2/18\\[1mm]
        1/18
\end{array}\right.
\quad\ \ \mbox{and}\quad\ \
\langle m_{\pi\pi} \rangle \, =\,
\left\{\begin{array}{ll}
        288/675\qquad \ \ \, & \mbox{for $RR$} \\[1mm]
        192/675\qquad \ \ \, & \mbox{for $RL/LR$}\\[1mm]
        128/675\qquad \ \  & \mbox{for $LL$}
\end{array}\right.
\end{eqnarray}
which are distinctly different for the pion final states. Due to the scalar stau
intermediate state, the $\tau \tau$ and $\pi \pi$ distributions do not depend
on the polarization state of the parent $\tilde{\chi}^0_2$ state [contrary to
the $q \tau$ and $q \pi$ distributions in the chain]. \s

Cutting out small transverse pion momenta in actual experiments
will modify the distributions of
the $\pi\pi$ invariant mass, and the distributions are shifted to larger
$m_{\pi\pi}$ values.
For $R$ chiralities of the $\tau$'s, with hard $\tau\to\pi$ fragmentation,
the shift is smaller than for $L$ chiralities with
soft fragmentation. [This will be analyzed quantitatively in the
next section for the specific reference point SPS1a.] \s\s

\begin{figure}[ht!]
\begin{center}
\includegraphics[height=11cm,width=15cm,angle=0]{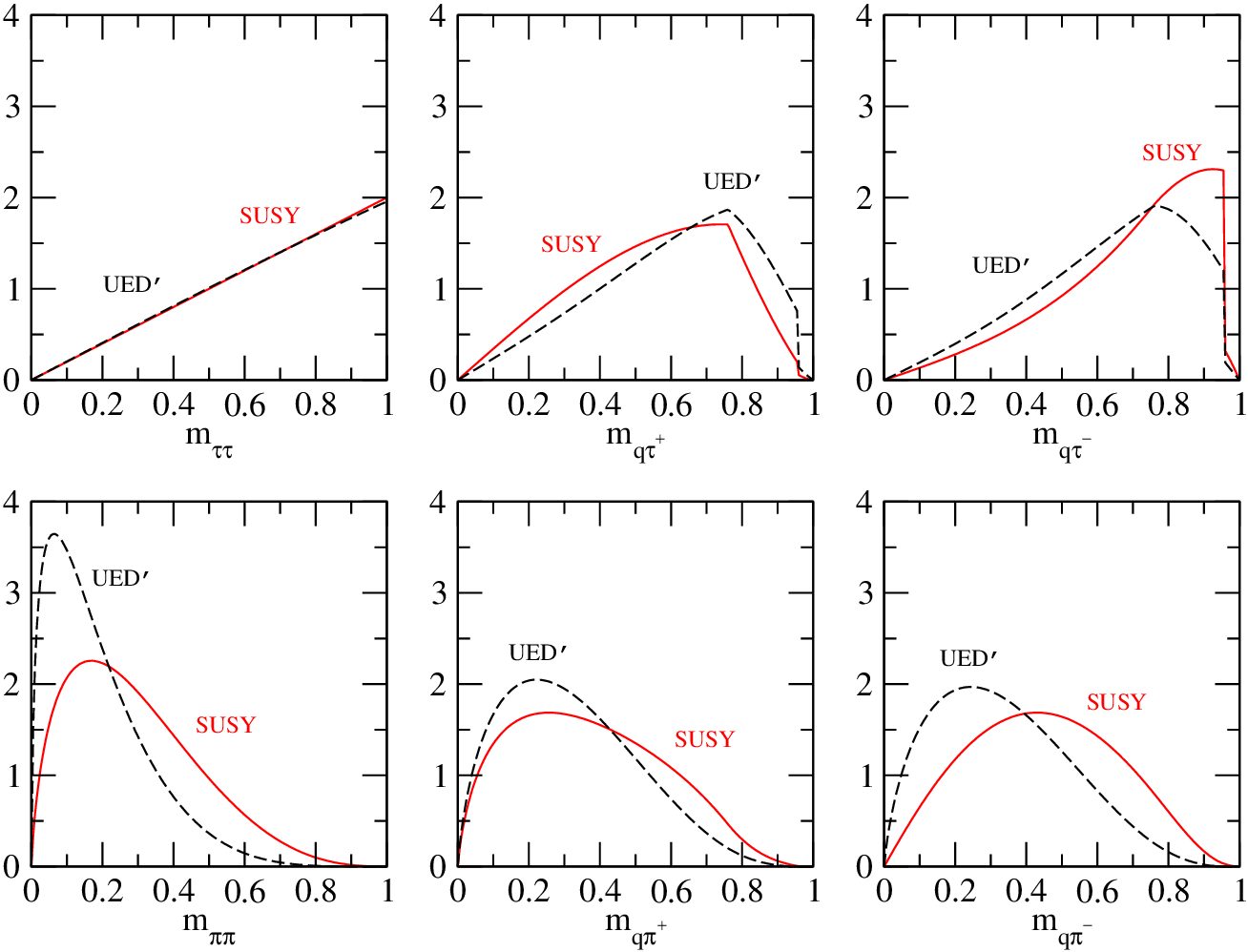}
\end{center}
\caption{\it The ``normalized"  distributions in supersymmetry
    and UED$'$ for the particle spectrum corresponding to the SUSY scenario SPS1a;
    the distributions $[\tau \tau],\, [q \tau^+]$ and $[q\tau^-]$ in the upper
    frames and the distributions $[\pi \pi],\, [q \pi^+]$ and $[q\pi^-]$ in
    the lower frames.}
\label{fig:sps1a_vs_ued}
\end{figure}
%


\noindent
{\bf 3.)} The sensitivity of various distributions $[\tau \tau],\, [q \tau^+],\,
[q \tau^-]$ and $[\pi \pi],\, [q \pi^+],\, [q \pi^-]$ in the cascade
(\ref{eq:susy_cascade_chain}) starting with a left-handed squark
$\tilde{q}_L$ is studied in the tableau
of Fig.$\,$\ref{fig:sps1a_vs_ued}. The distributions are compared for the SUSY
chain in SPS1a, in which the probability for $\tilde{\tau}_R$ in $\tilde{\tau}_1$
is close to $90\%$, with a spin chain (UED$'$) characteristic for
universal extra-dimension models
\cite{Appelquist:2000nn, Cheng:2002iz}:
\begin{eqnarray}
q_1\ \ \to\ \ qZ_1\ \ \to\ \ q\ell\ell_1\ \ \to\ \
q\ell\ell\gamma_1
\end{eqnarray}
To focus on the spin aspects,{\footnote{For this purpose we deliberately ignore
the naturally expected mass pattern in models of universal extra dimensions.}}
the same mass spectrum is chosen in the UED$'$ as
in the SUSY chain:
$m_{\tilde{q}_L}=m_{q_1}=570.6$ GeV, $m_{\tilde{\chi}^0_2}=m_{Z_1}=176.4$ GeV,
$m_{\tilde{\tau}_1}=m_{\tau_1}=134.1$ GeV and $m_{\tilde{\chi}^0_1}=m_{\gamma_1}
=98.6$ GeV. The $[\tau \tau]$ distribution is linear $\sim m_{\tau\tau}$ in
supersymmetry. It deviates slightly from the linear dependence in the $\tau\tau$
invariant mass in UED$'$ where in the limit of degenerate KK masses the
distribution reads $\sim m_{\tau\tau}[1 - \frac{1}{5}m^2_{\tau\tau}]$.
The kinks in the invariant mass distributions $[q \tau]$ and $[q \pi]$ signal
the transition from near to far $\tau$'s and $\pi$'s as the main components of
the events. Again, the best discriminant is the $[\pi \pi]$ distribution. This
reflects the $R$-dominated character of $\tilde{\tau}$ as opposed to the $L$ current
coupled to the Kaluza-Klein state $Z_1 \simeq W^3$ \cite{Cheng:2002iz}.\s

Experimental analyses of $\tau$ particles are a difficult task at LHC.
Isolation criteria of hadron and lepton tracks must be met which reduce the
efficiencies strongly for small transverse momenta.\footnote{We are very grateful
for comments on these problems by K.~Desch, D.~Mangeol and J.~Mnich.}
Stringent transverse momentum cuts
increase the efficiencies but reduce the primary event number and erase the difference
between $R$ and $L$ distributions. On the other hand, fairly small transverse momentum
cuts reduce the efficiencies but do not reduce the primary event number and the
$R/L$ sensitivity of the distributions.
Optimization procedures in this context are far beyond the scope of this
theoretical note. Experimental details may be studied in the recent
reports\cite{mangeol,unknown:1994pu} in which the analysis of di-$\tau$ final
states is presented for supersymmetry cascades of type (1) at LHC.\s

\begin{figure}[ht!]
\begin{center}
\includegraphics[height=10cm,width=16cm,angle=0]{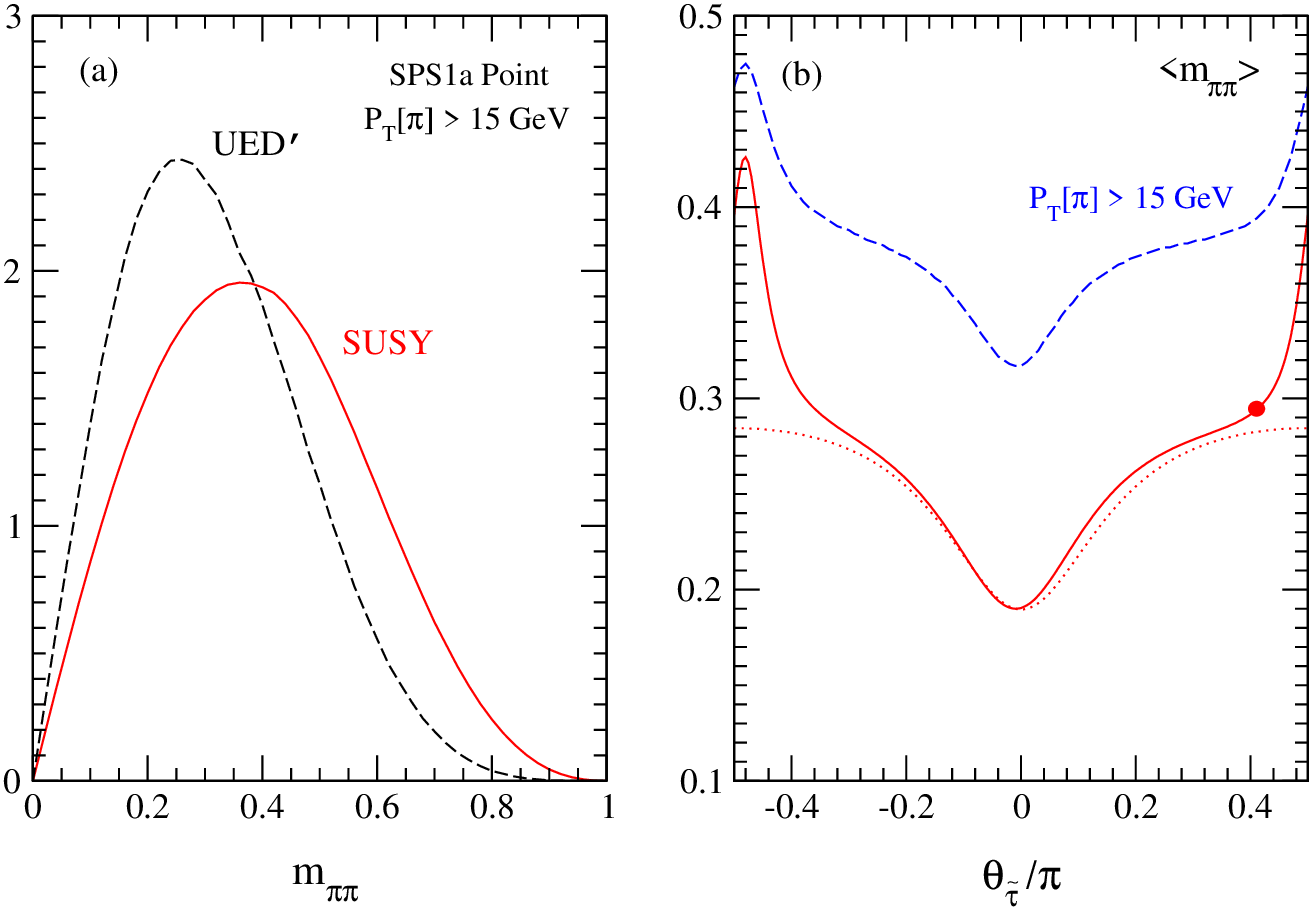}
\end{center}
\vskip -0.5cm
\caption{\it (a) The SUSY and UED$'$ distributions of the $\pi\pi$
   invariant mass with a cut of 15 GeV on the pion transverse momenta; and (b)
   the dependence of the expectation value $\langle m_{\pi\pi}\rangle$ on the
   stau mixing angle $\theta_{\tilde{\tau}}$ [in units of $\pi$] for the SPS1a values of
   $\tan\beta, M_{1,2}, \mu$ and the lighter stau mass $m_{\tilde{\tau}_1}$
   [without (red) and with (blue) $\pi$ transverse momentum cut].
   The (red) dot on the solid line denotes the value of $\langle m_{\pi\pi}\rangle$
   at the SPS1a point. The (red) dotted line represents the average values of
   $\langle m_{\pi\pi}\rangle$ if $\tilde{\chi}^0_{1,2}$ are unmixed
   pure bino-like and wino-like states, respectively.}
\label{fig:stau_mixing}
\end{figure}

In Fig.$\,$\ref{fig:stau_mixing}(a) it is shown how a cut of 15 GeV on the
pion transverse momenta modifies the SUSY and UED$'$ distributions of the
$\pi\pi$ invariant mass. Given the specific SPS1a mass differences, the
SUSY $LR$-dominated distribution is mildly affected while the UED$'$
$LL$-dominated distribution is shifted more strongly. [The different size
of the shifts can be traced back to the different shapes of the $R$ and $L$
fragmentation functions. Since $L$ fragmentation is soft, more events with
low transverse momentum are removed by the cut and the shift is
correspondingly larger than for hard $R$ fragmentation.]
Apparently, the transverse momentum cut does not erase the distinctive
difference between the $LR$ and $LL$ distributions. \s\s

%


\noindent
{\bf 4.)} Finally we study the dependence of the $[\pi\pi]$ distribution on
the stau mixing angle $\theta_{\tilde{\tau}}$, setting all other parameters
to their SPS1a values:
\begin{eqnarray}
&& \tilde{\tau}_1 \, =\, \phantom{-}\cos\theta_{\tilde{\tau}}\, \tilde{\tau}_L
                  +\sin\theta_{\tilde{\tau}}\, \tilde{\tau}_R \\
&& \tilde{\tau}_2 \, =\, -\sin\theta_{\tilde{\tau}}\, \tilde{\tau}_L
                  +\cos\theta_{\tilde{\tau}}\, \tilde{\tau}_R
\end{eqnarray}
The expectation value of the $\pi\pi$ mass distribution is given in terms of the
$\tau\tilde{\tau}\tilde{\chi}$ vertices in Eqs.$\,$(\ref{eq:tau_L_coupling}) and
(\ref{eq:tau_R_coupling}) by
\begin{eqnarray}
\langle m_{\pi\pi}\rangle &=& \frac{32}{675}\, \left(9\, \omega_{2R}\,\omega_{1R}
  +6\, \omega_{2R}\,\omega_{1L}+6\,\omega_{2L}\,\omega_{1R}
  +4\,\omega_{2L}\,\omega_{1L}\right)
\label{eq:pipi_average_formula}\\[2mm]
\omega_{j\alpha} &=& \frac{|A^\tau_{\alpha 1j}|^2}{
                       |A^\tau_{R1j}|^2+|A^\tau_{L1j}|^2} \qquad
                  [j=1,2; \alpha=R,L]
\end{eqnarray}
The coefficients are products of the average $\tau\tau$ invariant
mass and the average momenta in the $\tau \to \pi$ fragmentations.
The predictions for the $[\pi\pi]$ invariant-mass expectation value
are shown in Fig.$\,$\ref{fig:stau_mixing}(b). The effect of a
15 GeV cut on the transverse momenta of the pions is included for comparison.
The stau mixing angle of the SPS1a point, $\theta_{\tilde{\tau}}\approx 0.4\pi$,
is indicated by the red dot in the figure, representing a state
$\tilde{\tau}_1$ with 90\% $\tilde{\tau}_R$ and 10\% $\tilde{\tau}_L$ components.
The large sensitivity of the $[\pi\pi]$ invariant mass
to the mixing angle can be traced back to the fact that the neutralino
$\tilde{\chi}^0_2$ is nearly wino-like ($N_{22}=-0.94$). For
$\theta_{\tilde{\tau}} = \pm\pi/2$ the $R$ state $\tilde{\tau}_1
=\tilde{\tau}_R$, couples to $\tilde{\chi}^0_2$ only through its
small higgsino and U(1) gaugino components ({\it i.e.} $h_\tau N_{23}=0.04$ and
$N_{21}=-0.11$). However, once the mixing angle deviates slightly from $\pm \pi/2$
the ``near" tau lepton $\tau_{n}$ coupling quickly becomes $L$-dominated,
{\it cf.} Eqs.$\,$(\ref{eq:tau_L_coupling}/\ref{eq:tau_R_coupling}),
reducing $\omega_{RR}$ and $\omega_{RL}$ drastically and increasing
$\omega_{LR}$ significantly. [The sharp increase near the left edge is due
to the constructive interference between the U(1) gaugino and higgsino
contributions to the coupling $A^\tau_{R12}$.] The cut on the pion transverse
momenta modifies the $m_{\pi\pi}$ distribution, moderately for $R$ chirality
and strongly for $L$ chirality. Nevertheless, for each chirality combination
the shift is predicted completely in terms of the squark, $\tilde{\chi}^0_{1,2}$
and $\tilde{\tau}_1$ masses, and the shapes of the $R/L$ fragmentation functions,
so that it is under proper control. In particular, the relatively large increase
of the average $\langle m_{\pi\pi}\rangle$ near $\theta_{\tilde{\tau}}=0$
compared with $\pm\pi/2$ follows from the large effect of the transverse
momentum cut due to the soft $L$ fragmentation,
{\it cf.} Fig.$\,$\ref{fig:stau_mixing}(b).\s\s

\noindent
{\bf 5.)} {\it In summary.} The analysis of $\tau$ polarization in cascade decays
provides valuable information on chirality-type and mixing of supersymmetric
particles. The most exciting effects are predicted for the invariant mass
distributions in the $\pi\pi$ sector generated by the two polarized $\tau$
decays. This is strikingly different from the lepton-lepton invariant mass
distributions in the first two generations which do not depend on the
$R$ and $L$ couplings. It is particularly important to
notice that these polarization effects are independent of the couplings in the
squark/quark sector and also of the polarization state of the parent
$\tilde{\chi}^0_2$ generating the final $\tau \tau \tilde{\chi}^0_1$
state. \s\s

\subsection*{Acknowledgments}

Communications on experimental aspects of $\tau$ identification from
K.~Desch, D.~Mangeol and J.~Mnich are gratefully acknowledged. Also
discussions with M.~Nojiri have been of great help.
The work was supported in part by the Grant-in-Aid for Scientific
Research (No. 17540281) from the Japan Ministry of Education, Culture,
Sports, Science, and Technology. It was also
supported in part by the Korea Research Foundation
Grant (KRF-2006-013-C00097), by KOSEF through CHEP at Kyungpook
National University and by Deutsche Forschungsgemeinschaft.
S.Y.C. is grateful for support during his visit to DESY. P.M.Z.
thanks for the warm hospitality extended to him by the KEK Theory
Group.

\vspace{0.2cm}


\end{document}